\documentclass[preprint,12pt]{elsarticle}

\usepackage{amssymb}

\journal{Physics Letters A}

\begin{document}

\begin{frontmatter}

\title{Spin currents in non-inertial frames}

\author{Giorgio Papini}

\address{Department of Physics and Prairie Particle Physics Institute, University of Regina, Regina, Sask S4S 0A2, Canada}
\address{International Institute for Advanced Scientific Studies,
89019 Vietri sul Mare (SA), Italy}
\ead{papini@uregina.ca}
\begin{abstract}
The covariant Dirac equation and its solutions show that rotation and acceleration can be used to generate and control spin currents.

\end{abstract}

\begin{keyword}
Covariant Dirac equation, spin-gravity coupling, spin currents

PACS No.: 04.62.+v, 04.90.+e, 04.80.-y, 85.75.-d
\end{keyword}

\end{frontmatter}

{\it Introduction}. The realization that the flow of spin angular momentum can be separated from that of charge has
recently stimulated intense interest in fundamental spin physics \cite{ziese}. Spin control is an important
issue in spin based electronics \cite{igor} and in all phenomena essentially based on spin transport and dynamics \cite{bauer}.

The purpose of this work is
to study the generation and control of spin currents by rotation and acceleration.
In this context a fundamental tool is the covariant Dirac equation \cite{DE},\cite{14a}
\begin{equation}\label{CDE}
  \left[i\gamma^{\mu}(x)\left({\cal D}_\mu+ieA_{\mu}\right)-m\right]\Psi(x)=0\,,
  \end{equation}
which describes spin-$1/2$
fermions like electrons over wide energy ranges and is remarkably successful in dealing
with all inertial and gravitational effects discussed in the literature \cite{COW}-\cite{dinesh}.
The notations are those of \cite{pap2},
in particular ${\cal D}_\mu=\nabla_\mu+i\Gamma_\mu (x)$, $\nabla_\mu$ is
the covariant derivative, $\Gamma_{\mu}(x)$ the spin connection, commas indicate partial derivatives
and the matrices $\gamma^{\mu}(x)$ satisfy the relations
$\{\gamma^\mu(x), \gamma^\nu(x)\}=2g^{\mu\nu}$. In the absence of external fields (\ref{CDE}) reduces to the Dirac equation
\begin{equation}\label{DE}
\left(i\gamma^{\hat{\mu}}\partial_{\mu}-m\right)\psi_{0}(x)=0\,,
\end{equation}
where $\gamma^{\hat{\mu}}$ are the usual
constant Dirac matrices.

The covariant Dirac equation \cite{caipap3}, as well as other covariant wave equations \cite{papsc}-\cite{pap3}, can be solved exactly to first order
in the metric deviation $\gamma_{\mu\nu}= g_{\mu\nu}-\eta_{\mu\nu}$. This approximation is adequate for the
problems considered below.
The first
order solutions of (\ref{CDE}) are of the form
\begin{equation}\label{E}
  \Psi(x) = {\hat T}(x) \psi_{0}(x)\,,
\end{equation}
where $\psi_{0}(x)$ is a solution of (\ref{DE}), the operator $\hat{T}$ is given by \cite{pap2}
\begin{equation}\label{T}
    \hat{T}=
  -\frac{1}{2m}\left(-i\gamma^{\mu}(x)\mathcal{D}_{\mu}-m\right)e^{-i\Phi_{T}}\,,
\end{equation}
\begin{equation}\label{PHIS}
\Phi_{T}=\Phi_{S}+\Phi_{G}+\Phi_{EM},
\Phi_{S}(x)=\int_{P}^{x}dz^{\lambda}\Gamma_{\lambda}(z),\Phi_{EM}(x)=e\int_{P}^{x}dz^{\lambda}A_{\lambda}(z)
\end{equation}
and
 \begin{eqnarray}\label{PH}
  \Phi_{G}(x) = -\frac{1}{4}\int_P^xdz^\lambda\left[\gamma_{\alpha\lambda,
  \beta}(z)-\gamma_{\beta\lambda, \alpha}(z)\right]\left[\left(x^{\alpha}-
  z^{\alpha}\right)k^{\beta}-\left(x^{\beta}-z^{\beta}\right)k^{\alpha}\right]+
\\ \nonumber
 \frac{1}{2}\int_P^xdz^\lambda\gamma_{\alpha\lambda}(z)k^{\alpha}\,.
\end{eqnarray}
It is convenient to choose $\psi_{0}(x)$ in the form of plane waves, but wave packets can also be
used.

In (\ref{PHIS}) and (\ref{PH}), the path integrals are taken along
the classical world line of the particle, specifically, but not necessarily an electron, starting from a reference point
$P$. Only the path to $\mathcal{O}(\gamma_{\mu\nu})$ needs to be known in the integrations indicated
because (\ref{E}) already are first order solutions. The positive energy solutions of (\ref{DE}) are given by
\begin{equation}\label{psi0}
\psi(x)=u({\bf k})e^{-ik_\alpha x^\alpha}=N
  \left(\begin{array}{c}
                \phi \\
                 \frac{{\bf \sigma}\cdot{\bf k}}{E+m}\, \phi \end{array}\right)
                 \,e^{-ik_\alpha x^\alpha}\,,
\end{equation}
where $N=\sqrt{\frac{E+m}{2E}}$, $u^{+} u=1$, $\bar{u}=u^{+}\gamma^{0}$, $u^{+}_{1}u_{2}=u_{2}^{+}u_{1}=0$. In addition $\phi$ can take the forms $ \phi_{1}$  and $\phi_{2}$ where
$\phi_{1}=\left(\begin{array}{c}
1\\ 0 \end{array}\right)$, and $\phi_{2}=\left(\begin{array}{c}
0\\ 1 \end{array}\right)$.
When acceleration and rotation are present, $\gamma_{\mu\nu}$ is given by \cite{12},\cite{dinesh}
\begin{equation} \label{met}
\gamma_{00}\approx 2({\bf a}\cdot {\bf x})+({\bf a}\cdot {\bf x})^2-{\bf \Omega}^{2} {\bf x}^{2} +({\bf \Omega}\cdot {\bf x})^{2}\,, \\
\gamma_{0i}=-({\bf \Omega}\times {\bf x})_i\,,\gamma_{ij}=\eta_{ij}\,,
\end{equation}
where ${\bf a} $ and $ {\bf\Omega} $ represent acceleration and rotation respectively. To first order the tetrad is given by
\begin{eqnarray} \label{tet}
e^{\mu}_{\hat{\alpha}}&\approx &\delta^{\mu}_{\alpha}+h^{\mu}_{\hat{\alpha}}\,, h^{0}_{\hat{0}}=-{\bf a}\cdot {\bf x}\,,
h^{0}_{\hat{i}}=0\,, h^{k}_{\hat{i}}=0\,, h^{i}_{\hat{0}}=-\varepsilon^{ijk}\Omega_j x_k\,,\\
h^{\hat{0}}_{0}&=& {\bf a}\cdot {\bf x}\,, h^{\hat{k}}_{0}=\varepsilon_{ijk}\Omega^{i}x^{j}\,, h^{\hat{0}}_{i}=0, h^{\hat{k}}_{i}=\delta^{k}_{i} \nonumber \,.
\end{eqnarray}
from which the spinorial connection can be calculated using the relations
\begin{equation}\label{II.2}
  \gamma^\mu(x)=e^\mu_{\hat \alpha}(x) \gamma^{\hat
  \alpha}\,,\qquad
  \Gamma_\mu(x)=-\frac{1}{4} \sigma^{{\hat \alpha}{\hat \beta}}
  e^\nu_{\hat \alpha}\nabla_{\mu}e_{\nu\hat{\beta}}\,,
\end{equation}
where $\sigma^{{\hat \alpha}{\hat \beta}}=\frac{i}{2}[\gamma^{\hat
\alpha}, \gamma^{\hat \beta}]$. The result is $\Gamma_i =0$ and $\Gamma_0 =-\frac{1}{2}a_{i}\sigma^{\hat{0}\hat{i}}-\frac{1}{2}{\bf \Omega}\cdot{\bf \sigma}I$.

For electrons, $u_1$ corresponds to the choice $\phi =\phi_{1}$ and $u_2$ to $\phi =\phi_2$. Substituting in (\ref{psi0}), one finds
\[u_1=N
 \left(\begin{array}{c}
                1\\
                0\\
                \frac{k^3}{E+m}\\
                \frac{k^1 +ik^2}{E+m}\end{array}\right)\,,
                u_2=N
\left(\begin{array}{c}
                0\\
                1\\
                \frac{k^1-ik^2}{E+m}\\
                \frac{-k^3}{E+m}\end{array}\right)\,,\]
that are not eigenspinors of the matrix $\Sigma^3 =\sigma^3 I$ whose eigenvalues represent the spin components in
the $z$-direction, but become eigenspinors of $\Sigma^3$ when $k^1 =k^2 =0$, or in the rest frame of the electron $\textbf{k}=0$.

{\it Spin currents}. The transfer of angular momentum between the external non-inertial field and the electron spins can be demonstrated using the spin
current tensor \cite{grandy}
\begin{equation}\label{ST}
S^{\rho\mu\nu}=\frac{1}{4im}\left[\left(\nabla^{\rho}\bar{\Psi}\right)\sigma^{\mu\nu}(x)\Psi-\bar{\Psi}\sigma^{\mu\nu}(x)\left(\nabla^{\rho}\Psi\right)\right]\,,
\end{equation}
that satisfies the conservation law $S^{\rho\mu\nu},_{\rho}=0 $ when all $\gamma_{\alpha\beta}(x)$ vanish and yields in addition the expected result $S^{\rho\mu\nu}= \frac{1}{2}\bar{u}_{0}\sigma^{\hat{\mu}\hat{\nu}}u_{0}$ in the rest frame of the particle. Writing $\sigma^{\mu\nu}(x)\approx \sigma^{\hat{\mu}\hat{\nu}}+h^{\mu}_{\hat{\tau}}\sigma^{\hat{\tau}\hat{\nu}}+h^{\nu}_{\hat{\tau}}\sigma^{\hat{\mu}\hat{\tau}}$ and substituting (\ref{E}) and (\ref{T}) in (\ref{ST})
one obtains, to $\mathcal{O}(\gamma_{\alpha\beta})$,
\begin{equation}\label{STF}
S^{\rho\mu\nu}=\frac{1}{16im^3}\bar{u}_{0}\left\{8im^2k^{\rho}\sigma^{\hat{\mu}\hat{\nu}}+8imk^{\rho}h^{[\mu}_{\hat{\tau}}\sigma^{\hat{\tau}\hat{\nu}]}+ \right.
\end{equation}
\[
\left.
4imk^{\rho}\left(\Phi_{G,\alpha}
+k_{\sigma}h^{\sigma}_{\hat{\alpha}}\right)\left\{\sigma^{\hat{\mu}\hat{\nu}},\gamma^{\hat{\alpha}}\right\}-8imk^{\rho}\Phi_{G}k^{[\mu}\gamma^{\hat{\nu}]}+ \right.
\]
\[
\left.
4mk^{\rho}k_{\alpha}\left[\sigma^{\hat{\mu}\hat{\nu}},\left(\gamma^{\hat{\alpha}}\Phi_{S}-\gamma^{\hat{0}}\Phi^{+}_{S}
\gamma^{\hat{0}}\gamma^{\hat{\alpha}}\right)\right]+
4m^2k^{\rho}\left[\sigma^{\hat{\mu}\hat{\nu}},\left(\Phi_{S}-\gamma^{\hat{0}}\Phi^{+}_{S}\gamma^{\hat{0}}\right)\right]- \right.
\]
\[
\left. 8m^2 k^{\rho}h^{0}_{\hat{\alpha}}\left[\gamma^{\hat{0}},\left[\sigma^{\hat{0}\hat{\alpha}},\sigma^{\hat{\mu}\hat{\nu}}\right]\right] -8im^2k_{\sigma}\left(\Gamma^{\sigma}_{\alpha\beta}\eta^{\beta\rho}+\partial^{\rho}h^{\sigma}_{\hat{\alpha}}\right)\eta^{\alpha[\mu}\gamma^{\hat{\nu}]}+ \right.
\]
\[
\left. 8im^2\partial^{\rho}\Phi_{G}\left(4m\sigma^{\hat{\mu}\hat{\nu}}-2ik^{[\mu}\gamma^{\hat{\nu}]}\right)+4im^2\gamma^{\hat{0}}\Gamma^{\rho+}\gamma^{\hat{0}}\left\{\left(\gamma^{\hat{\alpha}}
k_{\alpha}+m\right),\sigma^{\hat{\mu}\hat{\nu}}\right\}\Gamma^{\rho} \right\}u_{0}
\]
where use has been made of the relation $ \Phi_{G,\mu\nu}=k^{\alpha}\Gamma^{\alpha}_{\mu\nu}$.
It is therefore possible to separate $S^{\rho\mu\nu}$ in inertial and non-inertial parts. The first term on the r.h.s. of (\ref{STF}) gives the usual result in the particle rest frame, when the external field vanishes. From (\ref{STF}) one finds
\begin{equation}\label{CST}
\partial_{\rho}S^{\rho\mu\nu}=\frac{1}{16im^3}\bar{u}_{0}\left\{8imk^{\rho}\partial_{\rho}h^{[\mu}_{\hat{\tau}}\sigma^{\hat{\tau}\hat{\nu}]}
-8imk^{\rho}\Phi_{G,\rho}k^{[\mu}\gamma^{\hat{\nu}]}+
\right.
\end{equation}
\[
\left.
4imk^{\rho}\left(k_{\sigma}\Gamma^{\sigma}_{\alpha\rho}+\partial_{\rho}h^{\sigma}_{\hat{\alpha}}k_{\sigma}\right)
\left\{\sigma^{\hat{\mu}\hat{\nu}},\gamma^{\hat{\alpha}}\right\}
+4mk^{\rho}k_{\alpha}\left[\sigma^{\hat{\mu}\hat{\nu}},\left(\gamma^{\hat{\alpha}}\Gamma_{\rho}-\gamma^{\hat{0}}\Gamma^{+}_{\rho}
\gamma^{\hat{0}}\gamma^{\hat{\alpha}}\right)\right]+ \right.
\]
\[
\left. 4m^2k^{\rho}\left[\sigma^{\hat{\mu}\hat{\nu}},\left(\Gamma_{\rho}-\gamma^{\hat{0}}\Gamma^{+}_{\rho}\gamma^{\hat{0}}\right)\right]+
8m^2 k^{\rho}\partial_{\rho}h^{0}_{\hat{\alpha}}\left[\gamma^{\hat{0}},\left[\sigma^{\hat{0}\hat{\alpha}},\sigma^{\hat{\mu}\hat{\nu}}\right]\right]- \right.
\]
\[
\left. 8im^2k_{\sigma}\partial^{\rho}\Gamma^{\sigma}_{\alpha\rho}\eta^{\alpha[\mu}\gamma^{\hat{\nu}]}+8im^2k_{\sigma}\Gamma^{\sigma}_{\rho\tau}\eta^{\tau\rho}
\left(4m\sigma^{\hat{\mu}\hat{\nu}}-2ik^{[\mu}\gamma^{\hat{\nu}]}\right)+ \right.
\]
\[
\left. 8im^2k^{\alpha}\Gamma^{\rho}_{\alpha\rho}\sigma^{\hat{\mu}\hat{\nu}}+8im^2k^{\rho}\Gamma^{\mu}_{\alpha\rho}\sigma^{\hat{\alpha}\hat{\nu}}+8im^2k^{\rho}\Gamma^{\nu}_{\alpha\rho}\right\}u_{0}\,, \]
where terms containing $\Gamma_{0,0}=0$ and $\partial_{\alpha}\partial_{\beta}h^{\mu}_{\hat{\nu}}=0$ have been eliminated. It follows from (\ref{CST}) that the external field invalidates the conservation law and that, therefore, there is continual interchange between spin and orbital angular momentum. The result is entirely similar to that observed for external electromagnetic fields \cite{grandy}. This essentially proves that non-inertial fields can be used in principle to {\it generate and control} spin currents. Because of the complexity of the equations, detailed studies of spin currents require lengthy calculations. In the rest frame
of the particle, when $ {\bf \Omega}=(0,0,\Omega)$, one finds
\begin{equation}\label{RF}
\partial_{\rho}S^{\rho\mu\nu}=\partial_{i}S^{i12}=\frac{1}{2}\left(\Gamma^{\rho}_{0\rho}+\Gamma^{1}_{10}+\Gamma^{2}_{20}\right)\bar{u}_{0}\sigma^{\hat{1}\hat{2}}u_{0}=\frac{E+m}{2E}\frac{\Omega a_{2} x}{1+{\bf a}\cdot {\bf x}}\,,
\end{equation}
and $ \partial_{i}S^{i13}= \partial_{i}S^{i23}=0$. In (\ref{RF}) $u_{0}$ corresponds to $u_{1}$. The violation of the conservation law is thus due to the direct coupling of the non-inertial field to the particle's spin current. Conservation is restored if either $\Omega$, or $a_{2}$,
or both vanish.

{\it Spin motion}. It is also useful to consider the actual spin motion. Transfer of angular momentum between external and non-inertial fields takes place when the operator $\hat{T}$
has some non-diagonal matrix elements. If in fact at time $t=0$ a beam of electrons is entirely of the $u_2$ variety, at time $t$ the fraction of
$u_1$ is $|\langle u_1|\hat{T}|u_2\rangle|^2$. It is more convenient to write the last expression in the form
\begin{equation}\label{me}
  P_{2\rightarrow 1}=\left|\langle u_1|\hat{T}|u_2\rangle\right|^{2}
  =\left|\int_{\lambda_0}^\lambda\langle u_1| {\dot x}^\mu {\hat
  T_{,\,\mu}}|u_2\rangle d\lambda\right|^2\\,
\end{equation}
 where $\dot{x}^{\mu}=k^{\mu}/m$ and $\lambda$ is an affine parameter along the electron world line. From \cite{pap2}
 \begin{equation}\label{tnu}
 \hat{T}_{,\,\nu} =\frac{1}{2m}\left\{h^{\mu}_{\hat{\alpha,\nu}}\gamma^{\hat{\alpha}}k_{\mu}+ \gamma^{\hat{\mu}}\Phi_{G,\mu\nu}-2im \left(\Phi_{G,\nu}+\Gamma_{\nu}-e A_{\nu}\right)\right\}\,,
  \end{equation}
  one can see that
 \begin{equation}\label{tnu}
 \langle u_1|\frac{k^{\nu}}{m}\hat{T}_{,\,\nu}|u_2\rangle =
 -i\frac{k^{0}}{m} \langle u_1|\Gamma_{0}|u_2\rangle=
 -i\frac{k^{0}}{m}\langle u_1|\left\{-\frac{1}{2} a_{i}\sigma^{\hat{0}\hat{i}}-
 \frac{1}{2}\Omega_{i} \sigma^{i} I \right\} |u_2 \rangle ,
 \end{equation}
 while the Mashhoon coupling $H_{M}=-{\bf \Omega}\cdot {\bf s}$ \cite{MASH}, where ${\bf s}=\frac{{\bf \sigma}}{2}$, and the interaction term $\frac{1}{2}a_{i}\sigma^{\hat{0}\hat{i}}=\frac{i}{2}a_{i}\alpha^{\hat{i}}$ give the first order equation of motion \cite{jackson}
 \begin{equation}\label{spin}
 \frac{d{\bf s}}{dt}={\bf s}\times \left({\bf \Omega}+ {\bf v}\times{\bf a}\right)\,,
 \end{equation}
which is useful in visualizing the spin motion under the action of rotation and acceleration. Notice that $A_{\mu}$ does not contribute to (\ref{tnu}) because $\langle u_1|u_2\rangle =0$ and that the terms $ie (h^{\mu}_{\hat{\alpha}}\gamma^{\hat{\alpha}}k_{\mu}+\gamma^{\hat{\mu}}\Phi_{G,\mu})A_{\nu}\frac{k^{\nu}}{m}$ drop out on account of $\langle u_1|\gamma^{\hat{\mu}}|u_2\rangle =0$.
No mixed effects of first order in rotation or acceleration and first order in the electromagnetic field are therefore present in this calculation. This applies to all terms containing the magnetic field ${\bf B}$, like the Zeeman term, and electric fields, like the spin-orbit interaction, that are present in the lowest order Dirac Hamiltonian that can be derived from (\ref{CDE}) \cite{dinesh}. To $\mathcal{O}(\gamma_{\mu\nu})$, contributions to (\ref{tnu}) from the electromagnetic field are present in the actual determination of the electron's path, as stated above.

From (\ref{tnu}) one obtains
\begin{eqnarray}\label{res}
 \frac{2m}{iE}\langle u_1|\frac{k^{\nu}}{m}\hat{T}_{,\,\nu}|u_2\rangle &=&-i\frac{k^3}{E}a_1-\frac{k^3}{E}a_2 +i\frac{k^1 -ik^2}{E}a_3\\ &+& \Omega^3 \frac{k^3}{E}\frac{-k^1 +ik^2}{E+m}\nonumber \\ &+&
 \Omega^1 \frac{E+m}{2E}\left(1+\frac{(k^{3})^2}{(E+m)^2}-\frac{(k^1 -ik^2)^2}{(E+m)^2}\right)\nonumber \\ &-&
 i\Omega^2 \frac{E+m}{2E}\left(1+\frac{(k^{3})^2}{(E+m)^2}+\frac{(k^1 -ik^2)^2}{(E+m)^2}\right)\equiv A_{12} \nonumber  \,,
 \end{eqnarray}
where $k^0\equiv E$. The parameter $k_{\mu}$ corresponds to the electron four-momentum when $ {\bf \Omega}=0$ and $ {\bf a}=0$.

Some general conclusions can be drawn from (\ref{res}). i) If $k^3 =0$, the particles move in the $(x,y)$-plane. If, in addition, $\Omega^1 =\Omega^2 =0$, then $A_{12}\neq 0$ only if $a_3 \neq 0$.
ii) If, however, ${\bf a}$ is also due to rotation, the conditions $\Omega_1 =\Omega_2 =0$ imply $a_3 =0$ and therefore $A_{12}=0$. This is the relevant case of motion in the $(x,y)$-plane with rotation along an axis perpendicular to it. One cannot therefore have a rotation induced spin current in this instance.
iii) Even for ${\bf k}=0$ one can have $A_{12}\neq 0$ if one of $\Omega_1$ and $\Omega_2$ does not vanish. This is a direct consequence of the spin rotation interaction or Mashhoon term contained in (\ref{CDE}) \cite{MASH},\cite{12},\cite{dinesh}.

A few examples are discussed below.

Consider an electron wave packet moving along the $x$-axis of a frame rotating about the same axis. Then ${\bf a}=0$ and the remaining parameters are
${\bf k}= (k,0,0)$ and ${\bf \Omega}=(\Omega,0,0)$. While the electron propagates along $x$, $u_1$ and $u_2$ propagate in opposite directions along
$x$ because of (\ref{tnu}) and (\ref{spin}). For a beam the spin current generated by rotation is therefore $I_s =I_{\downarrow}-I_{\uparrow}$, with obvious
meaning of the symbols. One finds $P_{2\rightarrow 1}=|\frac{i\Omega(E+m)}{4m}\int_{0}^{t}dt\left(1-\frac{k^2}{(E+m)^2}\right)|^2=(\frac{\Omega t}{2})^2$ which holds for $t\leq 2/\Omega$
on account of the requirement $P_{2\rightarrow 1}\leq 1$.

 Consider next a wave packet moving in  the plane $z=0$ itself
  moving along $z$ with velocity $k^3 /E$, while rotating about the $z$-axis with ${\bf \Omega}=(0,0,\Omega)$. The other parameters are ${\bf k}=(k\cos\Omega t, k\sin \Omega t,k^3)$ and ${\bf a}=(-\Omega^2 x,-\Omega^2 y,0)$. From (\ref{res}) one finds $A_{12}=(\frac{k^3}{E})(-ia^1-a^2 +\Omega \frac{-k^1 +ik^2}{E+m})$ and $P_{2\rightarrow 1}= \left[\frac{k^3}{E}\left(\Omega R+\frac{k}{E+m}\right)\sin2\Omega t\right]^2$, where $R$ is the radius of the circle described by the wave packet in the plane $z=0$. The motion of the center of mass of the wave packet is helical and so is the motion of the spin components which propagate, however, in opposite directions giving rise to a spin current.

In the case of motion occurring in the plane $z =0$ ($k^3=0$) and
${\bf \Omega}=(\Omega,0,0)$, one also gets ${\bf a} = (-\Omega^2 x,0,0)$ and
\begin{equation}\label{rot}
A_{12}=  \Omega \frac{E+m}{2E}\left\{1-\frac{(k^1-ik^2)^2}{(E+m)^2}\right\}\,,
\end{equation}
where $k^1 =k \cos\omega t$, $k^2 = k \sin \omega t$ and $\omega = \frac{eB}{m\gamma}$ is the cyclotron frequency of the electrons along the circular path determined by the constant magnetic field $B$. Substituting (\ref{rot}) in (\ref{me}) one finds
\begin{eqnarray}\label{rot1}
P_{2\rightarrow 1}&=&\left|i \Omega \frac{E+m}{2E}\int^{t}_{0}\left\{1-\frac{k^2 (\cos 2\omega t -i\sin 2\omega t)}{(E+m)^2}\right\}dt\right|^2 \\&=&
\left(\Omega t \frac{E+m}{4E}\right)^2 \left\{\frac{\sin^4 \omega t}{\omega^2 t^2}+\left(1-\frac{k^2}{(E+m)^2}\frac{\sin 2\omega t}{2\omega t}\right)^2\right\}\nonumber \,,
\end{eqnarray}
which holds for all $ t $ for which $P_{2 \rightarrow 1}\leq 1 $.
While both spin up and down electrons move on a circle of radius $R$ about the $z$-axis, they propagate in opposite directions because of the spin-rotation coupling, thus generating a spin current.

{\it Conclusions}. Rotation and acceleration can be used to generate and control spin currents. This follows from the covariant Dirac equation and its
exact solutions to $\mathcal{O}(\gamma_{\mu\nu})$.
To this order, external electromagnetic fields can be taken into account through the particle motion.
The transition amplitude for the conversion spin-up to spin-down is proportional to $A_{12}$ and is expressed as a
function of ${\bf \Omega}$, ${\bf a}$ and of the electron four-momentum $k_{\mu}$ before the onset of rotation and acceleration.
The same expression suggests criteria for the generation of spin currents and the transfer of momentum and
angular momentum to them. No energy can, of course, be transferred from the non-inertial fields to the spin currents as long as $\gamma_{\mu\nu}$ remains stationary.

\bibliographystyle{elsarticle-num}
\bibliography{<your-bib-database>}

\begin{thebibliography}{00}
\bibitem{ziese} M. Ziese and M. J. Thornton (Eds.), Spin Electronics, Springer-Verlag, Berlin, 2001.
\bibitem{igor} For a comprehensive review see: Igor \v{Z}uti\'{c}, Jaroslav Fabian, S. Das Sarma, Rev. Mod. Phys. {\bf 76}, 323 (2004).
\bibitem{bauer} Gerrit E. Bauer, Stefan Bretzel, Arne Brataas, Yaroslav Tserkovnyak, Phys. Rev. B {\bf 81}, 024427 (2010).
\bibitem{DE} C. G. De Oliveira and J. Tiomno, Nuovo Cimento
{\bf 24}, 672 (1962).
\bibitem{14a} A. Peres, Suppl. Nuovo Cimento {\bf 24}, 389 (1962).
\bibitem{COW} R. Colella, A. W. Overhauser and S. A. Werner, Phys. Rev. Lett. {\bf 34}, 1472 (1975).
\bibitem{PW} L. A. Page, Phys. Rev. Lett. {\bf 35}, 543 (1975).
             S. A. Werner, J-L. Staudenmann and R. Colella, Phys. Rev. Lett. {\bf 42}, 1103 (1979).
\bibitem{BW} V. Bonse and T. Wroblewski, Phys. Rev. Lett. {\bf 51}, 1401 (1983).
\bibitem{12} F. W. Hehl and W.-T. Ni. Phys. Rev. {\bf D42}, 2045 (1990).
\bibitem{dinesh} D. Singh, G. Papini, Nuovo Cim. B {\bf 115}, 223 (2000).
\bibitem{pap2} G. Lambiase, G. Papini, R. Punzi, G. Scarpetta, Phys. Rev. D {\bf 71}, 073011 (2005).
\bibitem{caipap3} Y. Q. Cai and G. Papini, Phys. Rev. Lett. {\bf 66}, 1259 (1991).
\bibitem{papsc}  G. Papini, G. Scarpetta, A. Feoli, G. Lambiase, Int. J. Mod. Phys. D {\bf 18}, 485 (2009).
\bibitem{pap1} G. Papini, Phys. Rev. D {\bf 75}, 044022 (2007).
\bibitem{pap3} G. Papini, Gen. Rel. Gravit. {\bf 40}, 1117 (2008).
\bibitem{grandy} Walter T. Grandy, Jr., Relativistic Quantum Mechanics of Leptons and Fields, Kluwer Academic Publishers, Dordrecht, 1990.
\bibitem{MASH} B. Mashhoon, Phys. Rev. Lett. {\bf 61}, 2639 (1988); Phys. Lett. A {\bf 139}, (1989); {\bf 143}, 176 (1990); {\bf 145}, 147 (1990); Phys. Rev. Lett. {\bf 68}, 3812 (1992).
\bibitem{jackson} J. D. Jackson, Classical Electrodynamics, Second Edition, John Wiley and Sons, New York, 1975.


\end{thebibliography}

\end{document}